\documentclass[12pt]{article} 

\usepackage{bbding}
\usepackage{mathtools}
\PassOptionsToPackage{normalem}{ulem}
\usepackage{ulem}

\usepackage{tikz,tikz-3dplot}
\usepackage{xcolor}
\usetikzlibrary{positioning}
\usetikzlibrary{3d}
\usepackage{cancel}
\usepackage{epsfig}
\usepackage{graphicx}
\usepackage{comment}
\usepackage{latexsym}
\usepackage{hyperref}
\usepackage{amsmath}
\usepackage{color}
\usepackage{amsbsy}
\usepackage{amssymb}
\usepackage{amsthm}
\usepackage{amsfonts}
\usepackage{cite}
\usepackage{enumitem}

\newcommand{\F}{{\cal F}}

\newcommand{\be}[0]{\begin{equation}}
\newcommand{\ee}[0]{\end{equation}}

\newcommand{\Z}{\mathbb{Z}}

\renewcommand{\and}{\mbox{and}}

\newcommand{\espD}{\phantom{\!\!\underset{\displaystyle |}{\cdot}}}
\newcommand{\espDD}{\phantom{\!\!\underset{\displaystyle |}{|}}}

\newcommand{\bm}{\boldmath} 
\catcode`\@=11
\def\marginnote#1{}
\newcount\hour
\newcount\minute
\newtoks\amorpm
\hour=\time\divide\hour by60 \minute=\time{\multiply\hour by60
\global\advance\minute by-\hour}
\edef\standardtime{{\ifnum\hour<12 \global\amorpm={am}%
        \else\global\amorpm={pm}\advance\hour by-12 \fi
        \ifnum\hour=0 \hour=12 \fi
        \number\hour:\ifnum\minute<10 0\fi\number\minute\the\amorpm}}
\edef\militarytime{\number\hour:\ifnum\minute<10 0\fi\number\minute}
\def\draftlabel#1{{\@bsphack\if@filesw {\let\thepage\relax
   \xdef\@gtempa{\write\@auxout{\string
      \newlabel{#1}{{\@currentlabel}{\thepage}}}}}\@gtempa
   \if@nobreak \ifvmode\nobreak\fi\fi\fi\@esphack}
        \gdef\@eqnlabel{#1}}
\def\@eqnlabel{}
\def\@vacuum{}
\def\draftmarginnote#1{\marginpar{\raggedright\scriptsize\tt#1}}
\def\draft{\oddsidemargin -.2truein
        \def\@oddfoot{\sl preliminary draft \hfil
        \rm\thepage\hfil\sl\today\quad\militarytime}
        \let\@evenfoot\@oddfoot \overfullrule 3pt
        \let\label=\draftlabel
        \let\marginnote=\draftmarginnote
   \def\@eqnnum{(\theequation)\rlap{\kern\marginparsep\tt\@eqnlabel}%
\global\let\@eqnlabel\@vacuum}  }
\def\thebibliography#1{
\vskip 0.5cm \centerline{\bf \Large References}
\list{
[\arabic{enumi}]}{\settowidth\labelwidth{[#1]}
\leftmargin\labelwidth
\advance\leftmargin\labelsep
\usecounter{enumi}}
\def\newblock{\hskip .11em plus .33em minus .07em}
\sloppy\clubpenalty4000\widowpenalty4000
\sfcode`\.=1000\relax}

\renewcommand{\theequation}{\arabic{section}.\arabic{equation}}
\renewcommand{\section}{\setcounter{equation}{0}\@startsection
{section}{1}{0mm}{-\baselineskip}{0.5\baselineskip} {\normalfont\Large\bfseries}}
\renewcommand{\subsection}{\@startsection
{subsection}{2}{0mm}{-\baselineskip}{0.5\baselineskip} {\normalfont\large\bfseries}}
\renewcommand{\subsubsection}{\@startsection
{subsubsection}{3}{0mm}{-\baselineskip}{0.5\baselineskip}
{\normalfont\normalsize\slshape}}

\begin{document}

%%%%%

\begin{titlepage}
\begin{flushright}
March 2020
\end{flushright}

\vspace{1cm}

\begin{centering}
{\bm\bf \Large Dark-matter spontaneous freeze out }\\

\vspace{12mm}
 {\bf Herv\'e Partouche}

\vspace{2mm}

{CPHT, CNRS, Ecole polytechnique, IP Paris, F-91128 Palaiseau, France\\ 
\vspace{0.3cm}
{\em herve.partouche@polytechnique.edu}} 
 \vspace{10mm}

{\bf\Large Abstract}

\end{centering}

%\vspace{4mm}

%\begin{quote}
\noindent 

{\noindent 

We consider the possibility that thermalized dark-matter particles acquire their mass thanks to the spontaneous breaking of a symmetry below some critical temperature. We describe the regime where a freeze out mechanism takes place shortly after the onset of the phase transition, while  the dark-matter mass has not yet reached its final constant value. For such a ``spontaneous freeze out''  to yield the correct relic density, the present-time cross section of annihilation of the dark matter into Standard-Model states has to be one or two orders of magnitude larger than in the case of a constant dark-matter mass.
}

\end{titlepage}
\newpage
\setcounter{footnote}{0}
\renewcommand{\thefootnote}{\arabic{footnote}}
 \setlength{\baselineskip}{.7cm} \setlength{\parskip}{.2cm}

\setcounter{section}{0}

%%%%%%%%%%%%%%%%%%%%%%%%%%%%%%%%

\section{Introduction}

%\section{Introduction}

In most models of cold dark matter (DM) production, the weakly interacting massive particle is considered to have a constant mass throughout the cosmological evolution. However, because all masses in the Standard Model (SM) arise through a Higgs mechanism, it may be more natural that a Higgs-like phase transition gives rise to the DM mass in a similar way. In this review, we will see that if a freeze out (FO) mechanism occurs during the phase where the DM mass evolves in time, the cross section of annihilation between DM and SM particles can be one or two orders of magnitude larger than in the constant DM-mass paradigm, in order to yield the correct relic abundance~\cite{SFO, SFO2,SFOstring}.  

\begin{figure}[h]
%\vspace{.4cm}
\begin{center}
\includegraphics[scale=0.7]{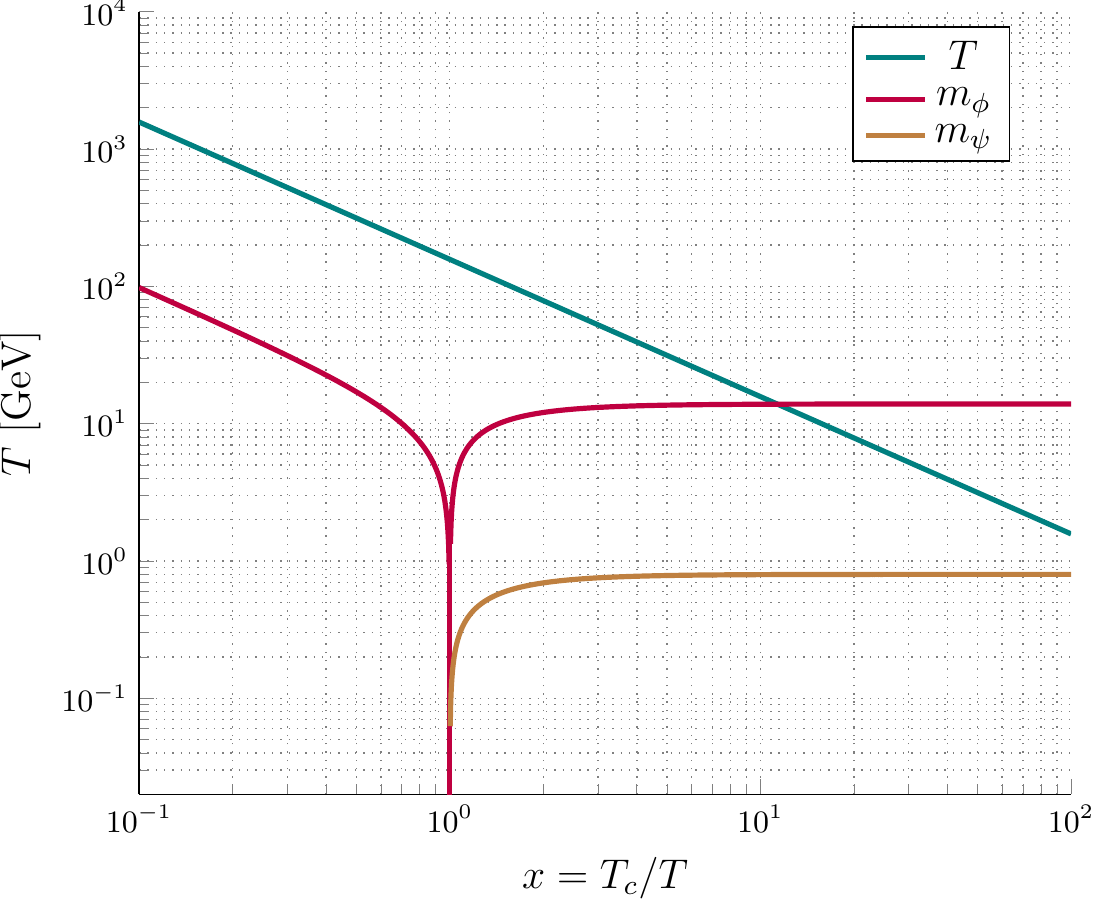}~~~~~~~
\includegraphics[scale=0.7]{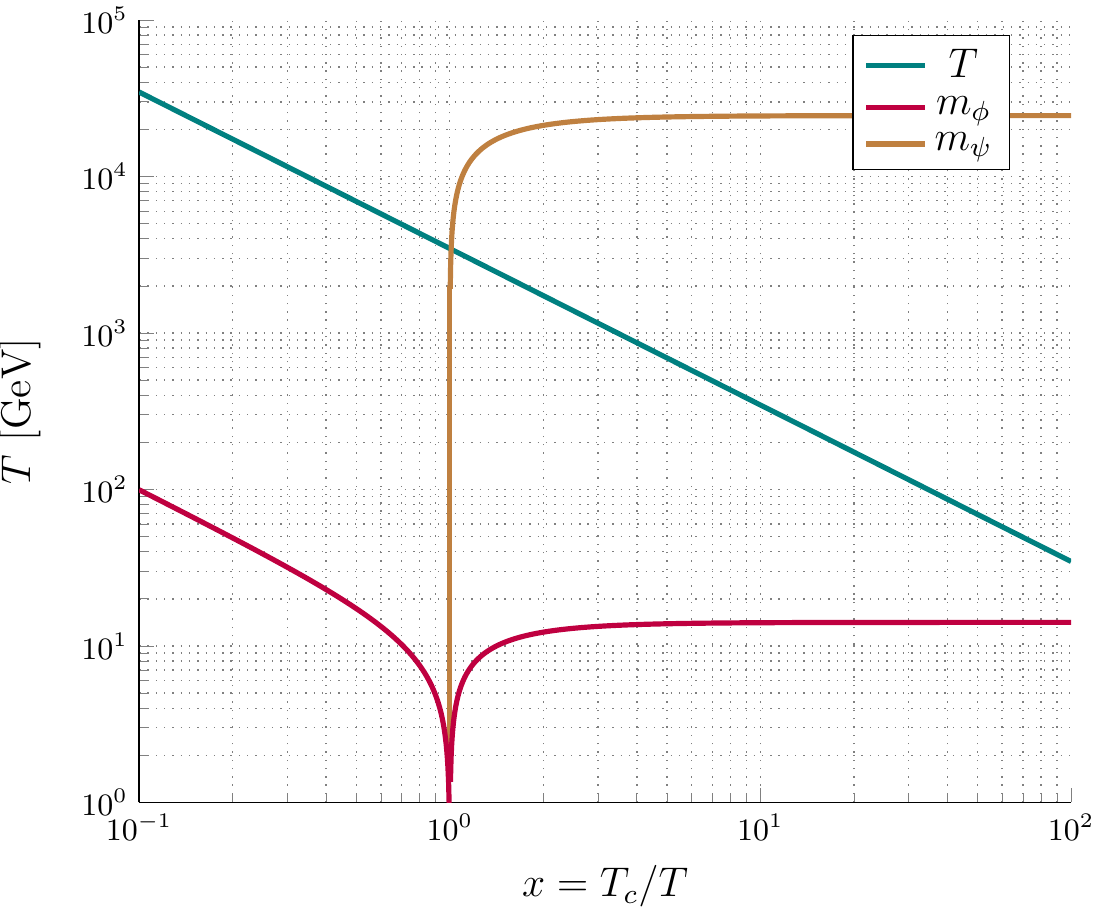}\end{center}
%\begin{picture}(0,0)
%\put(-36,97){$\V_{\rm th}$}
%\end{picture}
\caption{\em \footnotesize The temperature may reach $m_\psi$ when the latter is already constant (left panel). Alternatively, $m_\psi$ may exceed $T$ while it is still increasing (right panel).}
\label{FO-SFO}
\end{figure}

In the simplest realization of the mechanism, the dark sector includes a single Dirac or Majorana fermion $\psi$,  which will give rise to the relic density. Moreover, there is a real scalar $\phi$, which will not lead to some relic density, but will be responsible for the mass generation of $\psi$, thanks to a Yukawa interaction with coupling constant denoted $y$. At some early epoch after inflation, both $\psi$ and $\phi$ are assumed to be in thermal equilibrium with the SM. The vacuum expectation value (VEV) of $\phi$ is determined by the quantum effective potential at finite temperature $T$. The latter can yield two distinct phases characterized by a temperature higher or lower than a critical value $ T_c $, and giving rise to a trivial or non trivial VEV of $\phi$ and mass $m_\psi$ of $\psi$,
\be
\begin{aligned}
T>T_c~:&\quad  \langle \phi\rangle =0&&\Longrightarrow\quad m_\psi=0~,\\
T>T_c~:&\quad  \langle \phi\rangle \neq0\mbox{~~which depends on $T$}&&\Longrightarrow\quad m_\psi=y\langle \phi\rangle~.
\end{aligned}
\ee
Two qualitatively distinct regimes can be considered, as shown in   Fig.~\ref{FO-SFO}. The latter represents $T$  and the masses $m_\phi$, $m_\psi$ of $\phi$ and $\psi$ as functions of time, which is parametrized with a variable $x\equiv T_c/T$. 
As the universe expands, the temperature drops and may intersect $m_\psi$ when the latter has reached  its constant, asymptotic value (see left panel). When this occurs, a FO mechanism is expected to occur soon after, at a temperature $T^{\rm FO}$, as is the case is the standard constant-mass FO paradigm. In that case, we have $T_c\gg T^{\rm FO}$. Alternatively,  it may rather be $m_\psi$ that exceeds $T$, thanks to its rapid increase when the phase transition takes place. In that case, a FO mechanism, referred  to as ``Spontaneous Freeze Out'' (SFO),  is expected to occur at a temperature $T^{\rm SFO}$ close to $T_c$, when $m_\psi$ keeps on increasing (see right panel). We may compare the two scenarios  for a given mass  $m_\psi^0$ and yield $Y_\psi^0$ of the particles $\psi$ at present time. To this end, let us denote $m_\psi^{\rm FO}$ and $m_\psi^{\rm SFO}$ the masses of $\psi$ when the decoupling from the SM thermal bath takes place in both cases. By definition, we have $m_\psi^{SFO}<m_\psi^0=m_\psi^{FO}$. Hence, we expect the two FO temperatures to satisfy the same inequality, $T^{SFO}<T^{FO}$. This very fact implies that the interactions of $\psi$ with the SM  must be stronger in the SFO case, in order to maintain $\psi$ in thermal equilibrium at lower temperatures. As a result,  when the decouplings arise, the cross section of DM annihilation into  SM particles must be greater for the SFO case, as compared to that in the constant-mass case. Because the cross section increases with the mass, this conclusion is {\it a fortiori} true for cross sections measured at present time. 

At the classical level, the Lagrangian (in Dirac-fermion notation) can be decomposed as 
\be
\begin{aligned}
\mathcal L_{\rm tree}&=i\bar \psi\cancel{\partial}\psi+\frac{1}{2}(\partial \phi)^2-y \phi \bar\psi\psi -V_{\rm tree}(\phi)+\mathcal L_{\rm SM}+\mathcal L_{\rm int}~,\\ 
\mbox{where} \qquad V_{\rm tree}(\phi)&=-\frac{\mu^2}{2} \phi^2+\frac{\lambda}{4!}\phi^4~.
\end{aligned}
\ee
In these formulas, $\mathcal L_{\rm SM}$ stands for the SM Lagrangian, $\mathcal L_{\rm int}$ contains all interactions between the dark and observable sectors, and the potential of $\phi$, which depends on a tachyonic mass parameter $\mu$ and self-coupling $\lambda$,  is taken to be $\Z_2$ symmetric. Classically, the $\Z_2$ symmetry is  spontaneously broken, $\langle \phi\rangle\neq 0$. At one-loop, we add to the tree-level potential $V_{\rm tree}$ the contributions of the Coleman--Weinberg effective potential $V_{\rm CW}$ and of the free energy $\F$. Denoting $m_0^2\equiv -\mu^2+\lambda\phi^2/2$ and $m_\psi\equiv y\phi$, we may expand $\F$ at low $m_0^2/T^2$ and $m_\psi^2/T^2$, 
\be
\begin{aligned}
V_{\rm CW}(\phi)&=\frac{m_0^4}{64\pi^2} \!\!\left[\log\left(\frac{m_0^2}{Q^2}\right) -\frac{3}{2}\right]-n_F \frac{m_\psi^4}{64\pi^2} \left[ \log\left(\frac{m_\psi^2}{Q^2}\right) -\frac{3}{2}\right],\espDD\\
\F(T,\phi)&=-\frac{\pi^2}{90}T^4+\frac{T^2}{24}m_0^2 -{T\over 12\pi}\big(m_0^2\big)^{3\over 2} -\frac{m_0^4}{64\pi^2}\log \left(\frac{m_0^2}{16\alpha T^2}\right) +\cdots\espD\\
&~~~\,-n_F\Bigg[ \frac{7\pi^2}{720}T^4-\frac{T^2}{48}m_\psi^2-\frac{m_\psi^4}{64\pi^2} \log \left(\frac{m_\psi^2}{ \alpha T^2}\right)\!\Bigg]+\cdots~,  
\end{aligned}
\ee
where $n_F=4$ ($n_F=2$) for a Dirac (Majorana) fermion $\psi$,  $Q$ is a renormalization scale, and $\alpha=\pi^2e^{3/2-2\gamma_{\rm E}}$, with $\gamma_{\rm E}$ the Euler--Mascheroni constant. Notice that all logarithmic terms cancel out exactly, which avoids any imaginary contribution to arise from $\log m_0^2$ terms when $m_0^2<0$. However, the monomial $T(m_0^2)^{3/2}$ appearing in the free energy yields an imaginary contribution when $m_0^2<0$. To understand the origin of this pathology, observe that the $\Z_2$ symmetry is restored at high temperature, thanks to the thermal correction to the mass of $\phi$ arising from the term $T^2 m_0^2$. When this is the case, thermal loop corrections dominate the tree-level contribution, implying perturbation theory to break down. In order to be consistent at high temperature, one must therefore take into account additional thermal corrections occurring at arbitrary number of loops. The dominant ones arise from the so-called ``ring diagrams'' (see {\it e.g.} Ref.~\cite{Delaunay:2007wb} and references therein) and amout to adding
\be
\begin{aligned}
 V_{\rm ring}^{\rm th}(T,\phi)&={T\over 12\pi}\Big[ \big(m_0(\phi)^2\big)^{3\over 2} -\big(m_0(\phi)^2+\Pi_\phi(T)\big)^{3\over 2}\Big]~ ,\espD\\
 \mbox{where}\qquad  \Pi_\phi(T)&= {T^2\over 24}(\lambda+n_Fy^2)~.
\end{aligned}
\ee
As a result, the problematic term $T(m_0^2)^{3/2}$ is cancelled. Choosing $Q=\pi e^{-\gamma_{\rm E}} T_c$ and defining the critical temperature 
\be
T_c={2\sqrt{6}\, \mu\over \sqrt{\lambda+n_F y^2}}\, \sqrt{1-{\sqrt{6}\over 8\pi}\,\xi+{\log 2\over 8\pi^2}\,\lambda\over 1-{\sqrt{6}\over 4\pi}\,\xi}~,\qquad\mbox{where}\qquad  \xi\equiv {\lambda\over \sqrt{\lambda+n_Fy^2}}~, 
\ee
the total thermal effective potential may be approximated by  the expression 
\be
\begin{aligned}
V^{\rm th}_{\rm eff}(x,\phi)&=V_0(x)-\frac{\mu_{\rm eff}(x)^2}{2}\phi^2+\frac{\lambda_{\rm eff}(x)}{4!}\phi^4~,\espD\\
\mbox{where}\qquad\mu_{{\rm eff}}(x)^{2}&=\mu^2\bigg[\bigg(1-{\sqrt{6}\over 8\pi}\,\xi+{\log 2\over 8\pi^2}\,\lambda\bigg)\Big(1-{1\over x^2}\Big)-{\lambda\over 16\pi^2}\log x\bigg]\,,\espD\\
\lambda_{{\rm eff}}(x)&=\lambda\bigg(1-{3\sqrt{6}\over 8\pi}\,\xi+{3\log 2\over 8\pi^2}\,\lambda\bigg)+{3\over 16\pi^2}\big(4n_F y^4-\lambda^2\big)\log x~ .
\end{aligned}
\ee
As anticipated, $T_c$ deserves its denomination of critical temperature, thanks to the existence of two phases,
\be
 x<1 ~~ \Longrightarrow~~ \langle \phi\rangle = 0~, \qquad \quad x>1~~ \Longrightarrow~~ \langle \phi\rangle =  \mu_{{\rm eff}}(x)\sqrt{\frac{6}{\lambda_{\rm eff}(x)}}~.
\ee

The above formula at finite temperature are valid until $\psi$ or $\phi$ freezes out. In the following, we assume that $\psi$ is maintained in thermal equilibrium with the SM by contact interactions $\bar \psi\psi \bar ff$, where $f$ stands for SM fermions, and provided the cross section of annihilation $\psi+\bar \psi\to f+\bar f$ satisfies
\be
n_\psi\langle \sigma_{{\rm SM}\leftrightarrow\psi\bar\psi} \,v\rangle>H~, 
\ee
where $n_\psi$ is the number density of the particle $\psi$, $H$ is the Hubble parameter, $v$ is the relative velocity, and brackets signal a thermal average. Moreover, we assume that $\phi$ remains in thermal equilibrium, even after $\psi$ freezes out, thanks to interactions with the SM to be specified in the sequel. In order to characterize the region of parameter space where the FO mechanism is ``Spontaneous'', it is useful to define
\be
x_{\rm FO}\equiv{T_c\over T_{\rm FO}}~,  \qquad  \kappa={m_\psi(x_{\rm FO})\over T_{\rm FO}}~,
\ee
where $\kappa$ is in practice in the range $20\mbox{--}30$. Recall that in the constant-mass scenario, the ratio $x_{\rm FO}$ is very large.  From the above definitions, it is straightforward to show that 
\begin{align}
&\lambda\gg n_Fy^2~~ \Longrightarrow~~ x_{\rm FO}\simeq  {\cal O}\Big({2\kappa\over y}\Big)\gg \kappa~,  \\
&\lambda\ll n_Fy^2 ~~ \Longrightarrow ~~  1<x_{\rm FO}\simeq \left[1+\kappa^2\Big({4\lambda\over n_Fy^4}+{3\over \pi^2}\log x_{\rm FO}\Big)\right]^{1/2}<\kappa~,~~ \mbox{when}~~{4\lambda\over n_Fy^4}<1~.\nonumber 
\end{align}
Therefore, the SFO regime requires the last inequality on  the ratio  $4\lambda/( n_Fy^4)$ to hold. However, the latter cannot be arbitrarily small (and $x_{\rm FO}$ not too close to 1), in order for the effective potential at $T=0$ to admit a minimum at some $\langle \phi\rangle >0$. Taking into account this fact, one concludes that the SFO mechanism takes place when the equivalent following conditions are satisfied,  
\be
x_{\rm FO}\simeq \kappa ~~\quad \Longleftrightarrow~~\quad \frac{3 }{2\pi^2}\left(\log\frac{2}{3}+2\gamma_{\rm E}\right)\simeq 0.12<\frac{4\lambda}{n_Fy^4}<1~.
\ee

In order to check our expectations about the phenomenological features of the SFO scenario, let us assume that the Lagrangian $\mathcal L_{\rm int}$ contains an operator of the form
\be
\mathcal O_V = G_V\, \bar \psi {\gamma_\mu}\psi \bar f {\gamma^\mu}f\quad\mbox{ or }\quad \mathcal O_S = G_S\, \bar \psi\psi \bar f f~, 
\ee
where $G_V$, $G_S$ are coupling constants. These interactions lead respectively to $s$-wave  and $p$-wave thermally averaged annihilation cross sections, 
\be
\langle \sigma v\rangle_V \simeq  \frac{G_V^2}{2 \pi }\Big(1+\frac{x^{-1}T_c}{m_\psi(x)}\Big)m_\psi^2(x)~,\qquad 
\langle \sigma v\rangle_S\simeq  \frac{3G_S^2 }{8 \pi }\, x^{-1} T_c m_\psi(x)~,
\ee
that can be used to solve numerically the Boltzmann equation. The latter can be written as
\be
 \frac{{\rm d} Y_\psi}{\rm x}=\frac{\langle \sigma v\rangle s}{x H}(Y_{\psi,\rm{eq}}^{2}-Y_\psi^2)~,
\ee
where $Y_{\psi,\rm{eq}}$ denotes the yield of $\psi$ if it was in thermal equilibrium. We have scanned the parameter space of $\mu$, $\lambda$, $y$, $G_{V}$ or $G_S$, and selected the hypersurfaces that lead to the correct relic density \cite{SFO}. We find that in the SFO regime, {\it e.g.} when $x_{\rm FO}=1.1$, the mass of the DM particle increases by a factor of 2 between the time when the FO occurs and today. Moreover,  the cross sections at present time should 10 to 50 times larger than those needed when the DM mass is constant. Hence,  the SM fermion $f$ cannot be a quark, as it is already excluded  experimentally. On the contrary,  if $f$ is a lepton,  the vectorial operator $\mathcal{O}_V$ may  yield visible signals in the galaxy. Comparing  our simulations to the experimental limits set on DM annihilation by AMS \cite{Aguilar:2014mma,Accardo:2014lma} and Fermi \cite{Ackermann:2015zua, Fermi-LAT:2016uux}, we find that a DM candidate with a mass at the TeV scale could be detected in the near future by these experiments.

As already mentioned, in order to obtain these results, we have assumed that the scalar $\phi$ remains in thermal equilibrium with the SM. Before the phase transition at $T_c$, this can be assured by the (inverse) decay $\phi\leftrightarrow \psi+\bar \psi $. In the SFO regime, this process turns out to be  kinematically forbidden.  However, an effective decay of $\phi$ into the SM fermion $f$ exists, thanks to a loop of DM particles. One can also introduce a coupling between $\phi$ and  the SM Higgs boson. The reason why a decay of $\phi$ is important is that it increases the damping of the oscillations of $\phi$ in the well of its thermal effective potential, so that $\langle \phi\rangle$ actually tracks the $T$-dependent minimum. Indeed, without the effect of such a decay, when the temperature is below the mass of $\phi$, the energy stored in the oscillations of $\phi$ would contribute as the energy density of massive matter and overclose the universe.

%%%%%%%%%%%%%%%%%%%%%%%%%%%%%%%%%%

\bibliographystyle{unsrt}

\end{document}